# Restoration of tetragonal $C_4$ symmetry coexistent with filamentary superconductivity in the pressure induced intermediate phase in the iron-based superconductor $Ba_{1-x}K_xFe_2As_2$

Yan Zheng[1], Pok Man Tam[1], Jianqiang Hou[1], Anna E. Böhmer[2,3], Thomas Wolf[2], Christoph Meingast[2] and Rolf Lortz[1]♠

[1]*Department of Physics, The Hong Kong University of Science and Technology, Clear Water Bay, Kowloon, Hong Kong, China*
[2]*Institute for Solid State Physics, Karlsruhe Institute of Technology, PO Box 3640, 76021 Karlsruhe, Germany*
[3]*Fakultät für Physik, Karlsruhe Institute of Technology, 76131 Karlsruhe, Germany.*

The hole doped Fe-based superconductors $Ba_{1-x}A_xFe_2As_2$ (where A=Na or K) show a particular rich phase diagram. It was observed that an intermediate re-entrant tetragonal phase forms within the orthorhombic antiferromagnetically-ordered stripe-type spin density wave state above the superconducting transition [S. Avci *et al.*, Nature Comm. **5**, 3845 (2014), A. E. Böhmer *et al.*, arXiv:1412.7038v2]. A similar intermediate phase was reported to appear if pressure is applied to underdoped $Ba_{1-x}K_xFe_2As_2$ [E. Hassinger *et al.*, *Phys. Rev.* B **86**, 140502(R) (2012)]. Here we report data of the electric resistivity, Hall effect, specific heat, and the thermoelectrical Nernst and Seebeck coefficients measured on a $Ba_{0.85}K_{0.15}Fe_2As_2$ single crystal under pressure up to 5.5 GPa. The data reveals a coexistence of the intermediate phase with filamentary superconductivity. The Nernst coefficient shows a large signature of nematic order that coincides with the stripe-type spin density wave state up to optimal pressure. In the pressure-induced intermediate phase the nematic order is removed, thus confirming that its nature is a re-entrant tetragonal phase.

## I. INTRODUCTION

Applying external hydrostatic pressure to iron-based superconductors [1-6] has a similar effect as chemical ion substitution [7-13]. Especially in the $AEFe_2As_2$ ('122') materials (AE = Ba, Sr, Ca), the phase diagram is accessible through hole or electron doping or by application of external hydrostatic [2-6] and internal chemical pressure [7]. Electron doping can be achieved by substitution of $Fe^{2+}$ by $Co^{3+}$ [8], while hole doping occurs upon substitution of $Ba^{2+}$ by $K^+$ [9] or $Na^+$ [10]. Internal chemical pressure can be applied in the form of isovalent doping by introducing smaller ions in the structure, for example through substitution of As by P [13] or Fe by Ru [14]. Pressure is regarded generally as a particular clean way of tuning materials, since only one sample is used for the study of a particular region of the phase diagram, which normally helps to minimize reducing the effect of crystalline disorder [1,15]. The normal state of the 122 materials is characterized by both, a nematic order that reduces the rotational symmetry of the high-temperature tetragonal structure of $C_4$ symmetry to an orthorhombic $C_2$ symmetry, and a spin density wave (SDW) order that causes a stripe-type antiferromagnetic spin arrangement in the FeAs layers [16]. There is growing evidence that the nematic order is an electronically driven instability, but it remains unclear whether it is primarily the result of orbital order or spin driven order [17]. In doping the parent compound, e.g. $BaFe_2As_2$, the nematic and SDW ordering temperatures are gradually suppressed, and superconductivity appears with the maximum transition temperature $T_c$ close to the point where the transition lines meet the superconducting transition [16].

---

♠ Corresponding author: lortz@ust.hk

In hole-doped 122 compounds such as $Ba_{1-x}K_xFe_2As_2$ and $Ba_{1-x}Na_xFe_2As_2$ the phase diagram appears to be particular rich. In $Ba_{1-x}K_xFe_2As_2$, a new phase under pressure has been observed in resistivity measurements in the underdoped range with x = 0.16–0.19 [18], which appears within the magnetically ordered state. Only a small increase of $T_c$ from 16 K up to ~20 K was observed when pressure was applied to samples with x=0.18 up to 0.8 GPa. For higher pressure $T_c$ decreased slightly and then saturated at ~18 K up to at least 2.7 GPa. A second, step-like resistivity drop occurred above $T_c$, with its onset increasing to almost 50 K for pressure of 2.5 GPa. More recently, it was shown that such an intermediate phase also appears at ambient pressure in a narrow doping range between ~24% and 28% of K substitution, and was associated with a re-entrant tetragonal phase [19]. A similar ambient-pressure re-entrant $C_4$ phase was previously observed in $Ba_{1-x}Na_xFe_2As_2$ [20]. It is associated with a spin reorientation that restores the $C_4$ rotational symmetry [20,21], and has been interpreted as an itinerant double-Q spin density wave state [22]. It is still unclear whether the driving force this transition is due to magnetic interactions, or the orbital reconstruction of the iron $3d$ states [20]. The pressure-induced intermediate phase in $Ba_{1-x}K_xFe_2As_2$ [18] and the re-entrant $C_4$ phase at ambient pressure [19] are likely of similar origin.

In this paper we investigate the pressure phase diagram of a high quality [19] single crystal of $Ba_{0.85}K_{0.15}Fe_2As_2$ with a variety of experimental probes, including resistivity, Hall effect, specific heat and the thermoelectric Nernst and Seebeck coefficients, which we measure under pressure up to 5.5 GPa in a Bridgman type of pressure cell. This allows us to study the characteristics of the pressure-induced intermediate phase with a multitude of techniques that are rarely carried out under pressure. In particular the Nernst effect shows a strong sensitivity for the nematic order that coincides with the stripe-type SDW order up to optimal pressure. This nematic order is suppressed in the intermediate phase, which confirms that the rotational $C_4$ symmetry is restored in analogy to the re-entrant tetragonal phase at ambient pressure. In this pressure-induced re-entrant $C_4$ phase filamentary superconductivity exists up to 20 K above the bulk $T_c$.

## II. EXPERIMENTAL

The high pressure experiment was conducted in a modified Bridgman pressure cell in a pyrophillite gasket mounted on a tungsten carbide anvil with 3.5 mm active diameter [1,15]. As pressure medium, steatite was used, which has good quasi-hydrostatic conditions. Although not perfectly hydrostatic, its use is essential because of its low thermal conductivity that enables us to thermally isolate the samples from the anvils during the specific heat and thermoelectrical measurements. At temperatures exceeding 10 K this is much more difficult with most common liquid pressure media. The pressure gradients were always less than 0.1 GPa, as our manometer displayed. Although, small, the existence of some pressure gradient can alter the phase diagram of Fe-based superconductors in the vicinity of structural phase transition somewhat [4]. A photograph of the experimental set-up is shown in Fig. 1 together with a sketch showing the arrangement. Twelve 50 µm thin electric wires were fed through 70 µm narrow grooves in the gasket to electric contacts on the sample, to a resistive heater at a short distance from the sample, to two pairs of Au-Fe(0.07%)/Chromel thermocouples in contact with the samples and to electric contacts on a small piece of Pb foil, which serves as a manometer by monitoring the pressure dependence of its superconducting critical temperature [23-26]. The sample and the wires are placed on a thin disk made of steatite and fixed with tiny drops of epoxy resin. The Au wires are simply placed onto the sample and the manometer, so that the electric contacts will be established when pressure is applied to the cell. The contacts between the thermocouple legs were made with a small drop of silver epoxy. After completion of the set-up, it was covered by a second disk of steatite and some pyrophillite powder was placed to cover the wires in the grooves through the gasket for electric insulation from the top anvil.

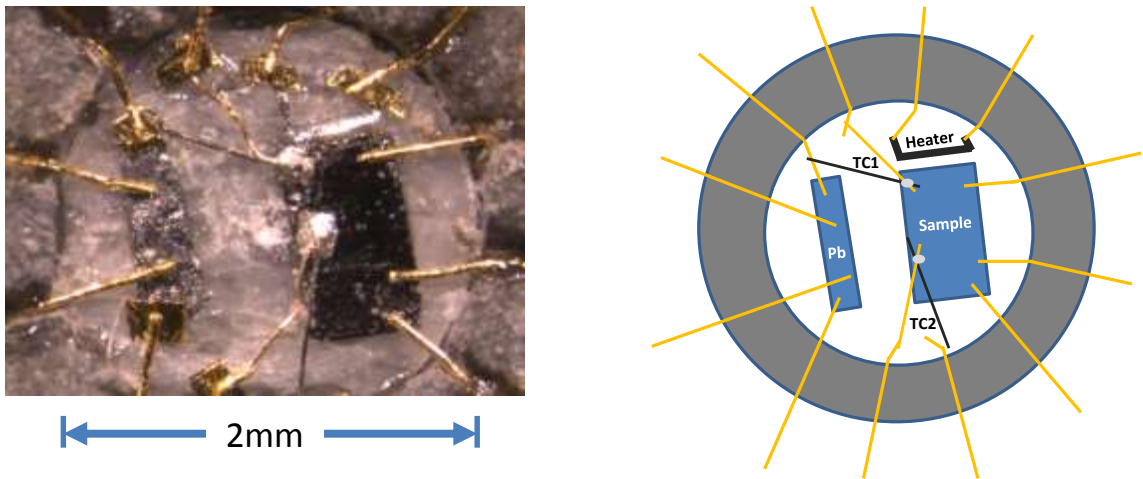

**Figure 1.** Experimental configuration in the gasket (gray ring on the edge of the picture) of the Bridgman cell on top of one anvil. (Left photograph, right sketch) The dark rectangle on the right is the $Ba_{0.85}K_{0.15}Fe_2As_2$ single crystal, contacted by Au leads and two thermocouples (TC1 & TC2). The light gray stripe on the left is a foil of Pb in a 4-probe resistance configuration that serves as manometer.

The electrical transport measurements were performed using a standard four-probe technique with an AC current source in combination with a lock-in technique. The specific heat was measured with an AC temperature-modulated technique at a high modulation frequency of several hundred Hz up to 1 kHz, to achieve thermal isolation of the sample [1,15,27]. The Joule-heating resistor was used to modulate periodically the temperature of the sample and a thermocouple was used to monitor the temperature modulation with help of a low-noise transformer connected to a lock-in amplifier. The thermoelectrical measurements were carried out in a similar way, however using transverse and longitudinal electric contacts on the sample to measure the Nernst and Seebeck voltages, respectively. The temperature gradient was determined by the two thermocouples. Both the Nernst effect and the Hall effect were measured for positive and negative magnetic fields applied perpendicular to the FeAs layers to eliminate spurious longitudinal voltages from an imperfect geometry of the electrodes. The Nernst effect data was measured during temperature sweeps at constant magnetic fields of ±6 T. The Hall effect was measured at stabilized temperatures during field sweeps.

## III. RESULTS

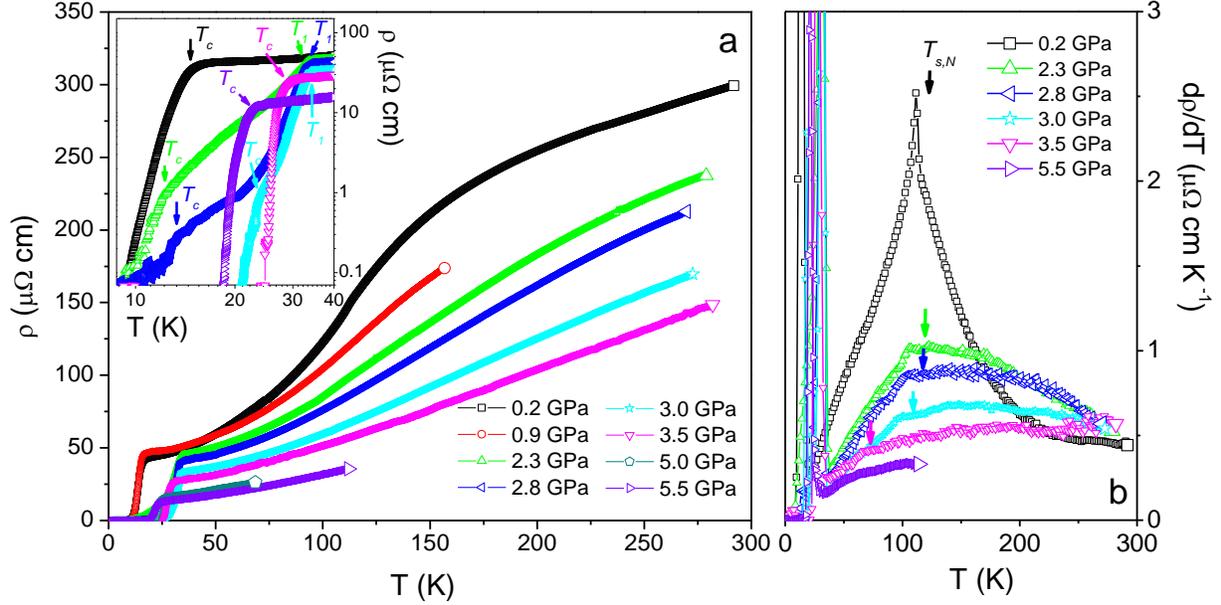

**Figure 2. a**: Resistivity of $Ba_{0.85}K_{0.15}Fe_2As_2$ under various pressures up to 5.5 GPa. The inset shows details at the superconducting transition in the low-temperature regime. **b**: Temperature derivative of the resistivity. The arrows mark the onset of nematic and stripe-type SDW order.

Fig. 2a shows the resistivity of the sample at different pressures up to 5.5 GPa. At the lowest pressure of $p = 0.24$ GPa, $T_c$ is indicated by a drop beginning at 14 K. Zero resistivity is reached at 10 K. A certain change in the slope marks the beginning of nematic and stripe-type SDW order below ~120 K. The transition is more obvious in the temperature derivative of the resistivity $d\rho/dT$ (Fig. 2b), which displays a sharp peak at $T_{s,N} = 112$ K. In higher pressure this magneto-structural transition is barely visible in the resistivity, which is likely a consequence of strain-induced broadening of the nematic transition by our quasi-hydrostatic conditions. However, the transition can be traced in form of a small bump up to 3.5 GPa in the derivative (Fig, 2b). At higher pressure, the onset of the resistive $T_c$ appears to increase rapidly and saturates at 34 K, but a tail of finite resistance remains down to 14 K and vanishes only at pressures of 3 GPa and higher. At higher pressure, $T_c$ decreases indicating that the over-doped regime is reached. At first glance, the overall behavior of $T_c(p)$ seems rather ordinary and the pressure-induced intermediate phase [18] is hidden in our experiment. The strong drop in the resistivity appears to be at first glance entirely related to superconductivity, with the $T_c$ onset increasing up to 34 K at 3 GPa. However, the long tail of the resistive transition extending down to ~14 K for pressures 1 - 3 GPa indicates that in this pressure range superconductivity is disturbed, which could be evidence for the presence of the intermediate phase. This is more evident in the inset in Fig. 2a, which shows the transition regime on a double logarithmic scale. In this plot, a double-step feature at $T_c$ (which we define in the following by the step-like anomaly in the low-temperature regime) and $T_1$ (defined at the onset of the upper step-like anomaly), becomes obvious for the data at 2.3, 2.8 and 3 GPa. This agrees with Hassinger's observations [18], although the drop in resistivity is much more pronounced in our experiment. A certain non-hydrostatic pressure component in our pressure cell likely causes some stress in the sample, and this may be the reason for the much lower resistivity of this intermediate phase in comparison to Hassinger's data [18]. In our experiment, the intermediate phase involves filamentary superconductivity, as we will demonstrate later.

In the following experiments, we concentrate on the low-pressure range up to 3 GPa, for which we have specific heat and thermoelectric data and investigate in detail the pressure-induced intermediate state between $T_c$ and $T_1$.

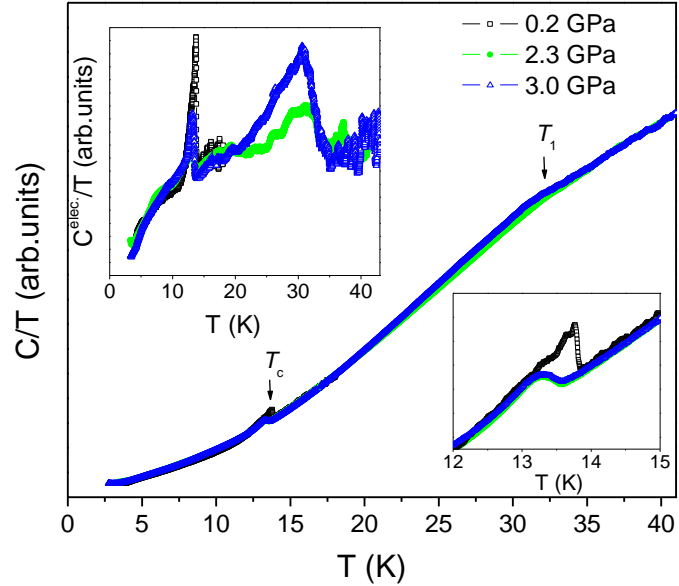

**Figure 3.** Specific heat of $Ba_{0.85}K_{0.15}Fe_2As_2$ under pressure of 0.2, 2.3 and 3.0 GPa. The lower right inset shows details at the superconducting transition temperature $T_c$. In the upper left inset, an approximate phonon background has been subtracted. This clearly reveals the presence of a further anomaly at $T_1 > T_c$ in the 2.3 and 3 GPa data.

Fig. 3 shows the specific heat at 0.2, 2.3 and 3 GPa. A sharp BCS-type superconducting transition with $T_c = 13.7$ K is seen at 0.2 GPa (see lower right inset for details). The specific heat jump occurs close to the temperature of the onset of the resistivity drop. At 2.3 and 3 GPa, $T_c$ does not increase, but actually decreases marginally to 13.5 K and remains at this temperature at 3 GPa and a further anomaly appears with maximum at ~32 K and onset at ~34 K for both pressures. In the upper left inset, we have subtracted an approximate phonon background, which shows the presence of the two distinct anomalies in this pressure range in detail. The onset of this transition anomaly coincides with the temperature where the resistance starts to drop in this pressure range. The specific heat as a bulk thermodynamic probe thus confirms that a distinct intermediate phase exists between $T_c = 13.5$ K and $T_1 = 34$ K for this pressure range, which is characterized by a small but finite resistance. The step-like shape of the anomaly at $T_1$ suggests that the transition is of second order, in contrast to the first-order transition that restores the $C_4$ rotational symmetry in $Ba_{1-x}K_xFe_2As_2$ at ambient pressure [19]. This may however be an artefact caused by pressure gradients.

The thermoelectrical Nernst effect has been proven to be a powerful tool for understanding the phase diagram of superconductors [28-30]. While the Nernst coefficient is usually very small for ordinary metals in their normal state, a large positive contribution from vortex motion results in type-II superconductors in addition to the regular normal state contribution, and has been used to monitor the presence of superconducting fluctuations above $T_c$ [28]. In addition, the Nernst coefficient is particularly sensitive to nematic order arising from electronic correlations that spontaneously break the rotational symmetry, as observed in various cuprate and Fe-based superconductors [30].

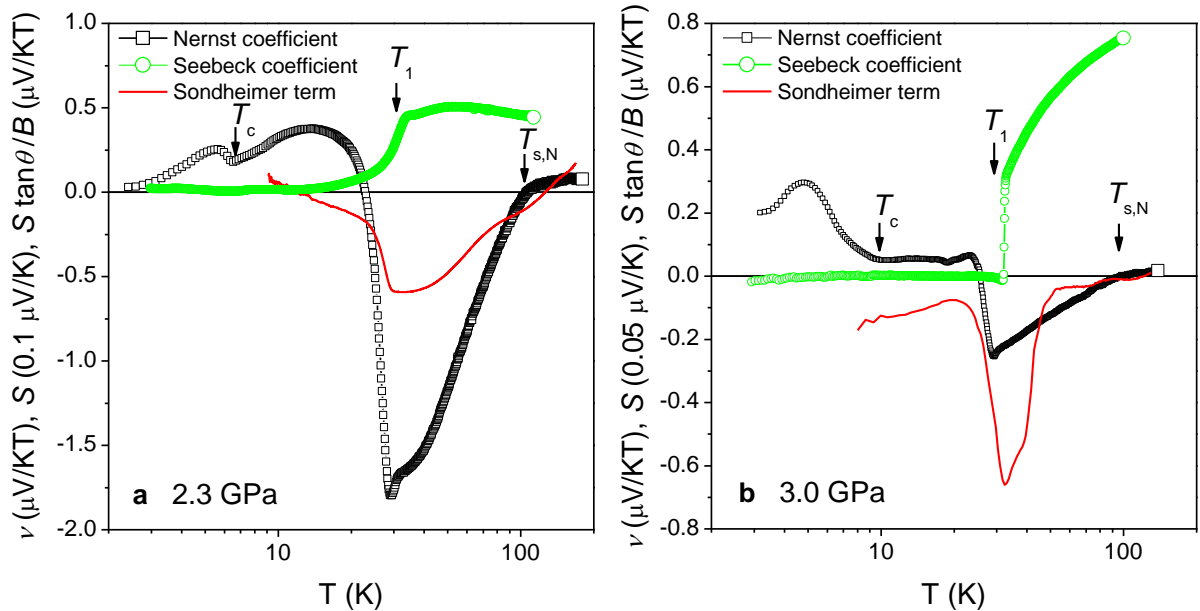

**Figure 4.** Temperature dependence of the Nernst coefficient $\nu = N/B$ in a field of 6 T applied perpendicular to the FeAs layers (black open dots), the zero-field thermopower (green squares, scaled for comparison) and the Sondheimer term $S\tan\theta/B$ (red line) of $Ba_{0.85}K_{0.15}Fe_2As_2$ at pressure of 2.3 GPa (**a**) and 3 GPa (**b**).

Fig. 4 shows the thermoelectric Nernst (in a 6 T field) and zero-field Seebeck coefficients measured at 2.3 and 3 GPa. The Seebeck coefficient primarily mimics the resistivity and drops at both pressures below ~34 K, presumably due to filamentary superconductivity. The Nernst coefficient shows two distinct anomalies, which we attribute to the transitions observed at $T_c$ and $T_1$ in the specific heat. The fact that the anomaly at $T_c$ is suppressed to below 10 K can be explained by the applied 6 T magnetic field. A pronounced positive signal appears for both pressures below $T_c$, which clearly corresponds to vortex motion [28]. The data remains positive up to $T_1$, although the data in 2.3 and 3 GPa show a somewhat different behavior in this temperature regime. At 2.3 GPa, there is a broad positive bump between $T_c$ and $T_1$. At 3 GPa, a plateau with finite small positive value is observed in this temperature range. At $T_1$ the Nernst coefficient changes abruptly to a large negative signature in the form of a steep drop when the temperature is increased. Such a giant normal-state Nernst coefficient is regarded as a signature of nematic order, which is related to a symmetry-breaking Fermi surface reconstruction [30]. It has been observed previously for $LaFeAsO_{1-x}F_x$ [29] and $CaFe_{2-x}Co_xAs_2$ [32], but its magnitude is particularly large here. In the high temperature region, the negative signal magnitude decreases gradually and crosses zero at ~90 K before it saturates at a small positive value at ~104 K where the nematic and SDW order vanishes. The magnitude of this negative contribution is smaller at 3

GPa, which indicates that the nematic and SDW orders become suppressed rapidly at higher pressure, even though the antiferromagnetic transition appears only slightly decreased to ~94 K. The Nernst data further confirms that a distinct intermediate phase is present between $T_c$ and $T_1$, which shows similar characteristics to the re-entrant $C_4$ phases observed in $Ba_{1-x}K_xFe_2As_2$ [19] and $Ba_{1-x}Na_xFe_2As_2$ [20] at ambient pressure. The vanishing of the strongly negative Nernst coefficient in the SDW phase below $T_1$ demonstrates that the nematic order is suppressed in this phase. Its finite positive Nernst coefficient points to the presence of mobile vortices well above the bulk superconducting transition at $T_c$. We can thus conclude that the intermediate phase is a re-entrant $C_4$ phase like the one observed at ambient pressure [19] and coexists in our experiment with filamentary superconductivity.

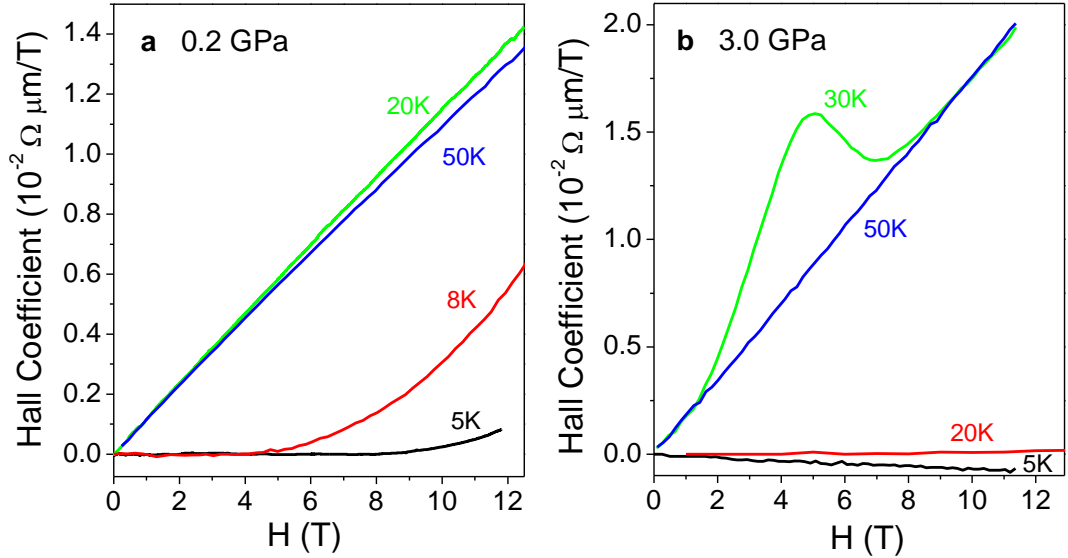

**Figure 5.** Field dependence of the Hall coefficient at various temperatures in 0.2 GPa (a) and 3 GPa (b).

Fig. 5 shows the Hall coefficient at various fixed temperatures. The magnetic field was applied perpendicular to the FeAs layers. Here we compare the behavior at very low pressure of 0.2 GPa without intermediate phase with data at 3 GPa with the intermediate phase. The Hall voltage vanishes in the entire temperature range below the bulk superconducting $T_c$. In the normal state it shows the expected linear field dependence. At 0.2 GPa and 5 K, the Hall voltage is zero in field below a characteristic critical field $H_c \sim 9$ T, and then increases gradually. At 8 K, the trend is similar, but with a lower $H_c \sim 4.5$ T. At 3 GPa, the Hall coefficient almost vanishes at 20 K, and obviously $H_c$ exceeds our available field range up at this pressure. The vanishing Hall coefficient confirms that the material is still in a superconducting state from a transport point of view, in accordance with our conclusions from the Nernst coefficient and the resistivity. At 5 K, a slightly negative Hall signal is observed, presumably related to vortex flow. It is interesting that at 30 K the signal increases with a larger slope first and then goes through a maximum and finally the linear normal-state behavior is approached. This characteristic is absent at higher temperatures. It is obviously related to the vicinity of the transition from the re-entrant $C_4$ phase to the stripe-type SDW phase and may be a precursor of the Fermi surface reconstruction associated with the appearance of the stripe-type SDW and nematic order [31].

## IV. DISCUSSION

Under the influence of a temperature gradient and an electrical field the total charge carrier current density is $J = \bar{\bar{\sigma}} \bullet E + \bar{\bar{\alpha}} \bullet (-\nabla T)$, where $\alpha$ is the Peltier coefficient tensor and $\sigma$ is the conductivity tensor. The resulting transverse Nernst voltage is then:

$$N = \frac{E_y}{\nabla_x T} = \frac{\alpha_{xy}\sigma_{xx} - \alpha_{xx}\sigma_{xy}}{\sigma_{xx}^2 + \sigma_{xy}^2} \tag{4}$$

The solution of the Boltzmann equation leads to the following relationship between the electrical and the thermoelectrical conductivity tensors:

$$\bar{\alpha} = -\frac{\pi^2}{3}\frac{k_B^2 T}{e}\frac{\partial \sigma}{\partial \varepsilon}\bigg|_{\varepsilon=\varepsilon_F} \tag{5}$$

Substitution into equation 4 yields:

$$N = -\frac{\pi^2}{3}\frac{k_B^2 T}{e}\frac{\partial \tan \theta_H}{\partial \varepsilon}\bigg|_{\varepsilon=\varepsilon_F} \tag{6}$$

For a single band system, the Hall angle can be expressed in terms of the cyclotron frequency and scattering time. Therefore, an alternative description of the Nernst effect is [33]:

$$v = N/B = -\frac{\pi^2}{3}\frac{k_B^2 T}{m^*}\frac{\partial \tau}{\partial \varepsilon}\bigg|_{\varepsilon=\varepsilon_F} \tag{7}$$

where $B$ is the magnetic field. Obviously, the Nernst signal is zero when the Hall angle is independent of energy. Taking into account equation (6) for the quasi-particle contribution only, it is clear that two cases lead to observable Nernst signals: a multi-band structure or energy-dependent Hall angles. The characteristic Fermi surface of Fe-based superconductors is responsible for the occurrence of charge carriers of two opposite signs [16] due to multiple bands. Therefore, it is crucial to analyze the quasiparticle contribution to the Nernst signal. To what extend the Nernst signal is unusual can be judged from the 'Sondheimer cancellation' [33]

$$v = \left(\frac{\alpha_{xy}}{\sigma} - S\tan\theta\right)\frac{1}{B},$$

where $S$ is the Seebeck coefficient and $\theta$ the Hall angle. In an ordinary one-band metal the two terms cancel, resulting in a vanishing of the Nernst coefficient $v$. To test whether the Sondheimer cancellation holds is therefore a powerful tool to reveal strongly correlated electronic states such as superconductivity, nematic order or SDW states. The degree of violation of the Sondheimer cancellation can be tested experimentally, by comparing the measured $v$ with the term $S\tan\theta/B$ derived from thermopower and Hall effect data. For an ordinary metal it would be expected that $v$ is much smaller than $S\tan\theta/B$.

In Fig. 4 we added the Sondheimer terms to the plot of the Nernst and Seebeck coefficients to illustrate the result of the Sondheimer cancellation. At 2.3 GPa, the absolute magnitude of $v$ in the stripe-type SDW and nematic state is much larger than $S\tan\theta/B$. For LaFeAsO, the violation of the Sondheimer cancellation in this state has been explained by a Fermi surface reconstruction [29] with spontaneous breaking of rotational symmetry associated with the nematic order [31]. Upon comparing the magnitude of the negative Nernst signal with the parent compounds LaFeAsO [29] and $CaFe_2As_2$ [32], it becomes obvious that the absolute magnitude of the Nernst coefficient associated with the nematic order is particularly large for $Ba_{1-x}K_xFe_2As_2$ in this pressure range, especially regarding the fact that optimal pressure is almost reached.

The absolute magnitude of the negative signal in the stripe-type SDW state above $T_1$ is much larger than that of the positive vortex contribution in the bulk superconducting state below $T_c$. Therefore, the positive Nernst coefficient in the intermediate phase between $T_c$ and $T_1$ cannot originate from a partial cancellation of positive and negative contributions. The negative Nernst signal of the nematic order thus vanishes below $T_1$ and becomes replaced by a positive weak vortex contribution of superconducting origin. The vortex signal is weaker than in the bulk superconducting state and the resistivity is finite, thus identifying a filamentary nature of superconductivity. A likely explanation is that the transition at $T_1$ into the re-entrant $C_4$ phase is incomplete and some orthorhombic domains coexist with tetragonal domains in the volume. The pressure medium in our experiment is less hydrostatic compared to Hassinger's conditions [18]. It was reported that the transition from the nematic SDW phase into a re-entrant $C_4$ phase in $Ba_{1-x}Na_xFe_2As_2$ is not always complete and some phase separation with spatially separated orthorhombic and tetragonal regions may exist [21]. This may depend significantly on the choice of pressure medium. Our data does not supply any information about the structure or magnetic order in the intermediate phase, but a likely explanation for our data is coexistence between a re-entrant $C_4$ phase and a $C_2$ minority phase on a microscopic length scale. The filamentary superconductivity could be associated either with the $C_2$ minority phase, or the domain boundaries between the two phases. On the other hand, the volume fraction occupied by the orthorhombic domains must be very small, since the Nernst coefficient does not show any sign of long range nematic order in the re-entrant $C_4$ phase.

At 3 GPa, the sample approaches optimal pressure conditions and the negative Nernst contribution in the stripe-type SDW phase is much smaller compared to the sample at 2.3 GPa. Indeed, the Nernst coefficient $v$ is now smaller than $S \tan\theta/B$ and thus the Sondheimer cancellation is no longer violated. This is in accordance with the expectation that the nematic order vanishes upon approaching the overdoped regime.

## V. CONCLUDING REMARKS

Our electric transport, specific heat and thermoelectric data confirms the presence of a pressure-induced intermediate phase between the onset of bulk superconductivity and the stripe-type SDW state observed by Hassinger *et al.* [18]. In addition, our Nernst effect data shows that the nematic order is absent in this phase, thus confirming that it represents a similar re-entrant $C_4$ phase as observed for $Ba_{1-x}K_xFe_2As_2$ with higher K contents at ambient pressure. The data further reveals a highly complex interplay of superconductivity, SDW and nematic order. A compiled phase diagram is shown in Fig. 6. The new phase is referred to as *"Filamentary SC + re-entrant tet."*. In this intermediate phase, tetragonal regions likely coexist spatially separated with orthorhombic regions of vanishingly small volume fraction in which the ordinary stripe-type antiferromagnetic SDW order is preserved. This coexistence causes filamentary superconductivity. When the temperature is lowered, the entire sample goes into a bulk superconducting state in which the larger positive Nernst coefficient indicates a liquid vortex phase. At higher pressure, the intermediate state shrinks and finally disappears at 3 GPa. We added Hassinger's data [18] for a 16% K doped sample (stars) in the phase diagram, which is in perfect agreement with our data.

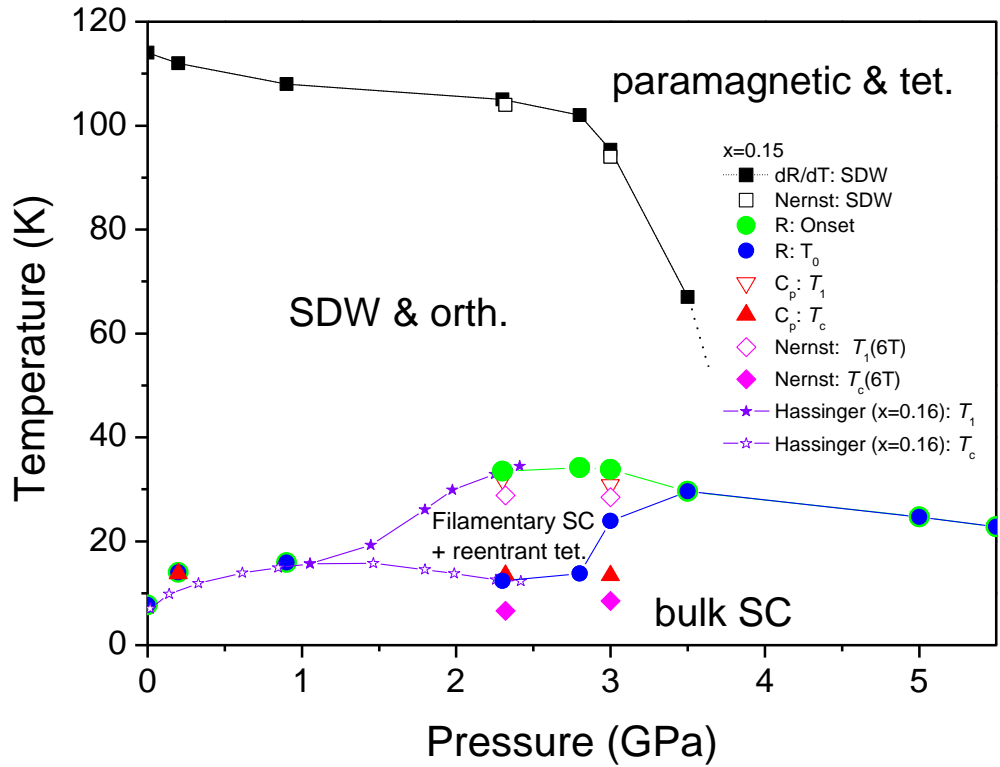

**Figure 6.** Phase diagram of $Ba_{0.85}K_{0.15}Fe_2As_2$ compiled from the resistivity, specific heat and Nernst coefficient data (tet.: tetragonal, orth.: orthorhombic, SDW: stripe-type antiferromagnetic spin density wave state, SC: superconducting state). The stars are data taken from Ref. 16 for $Ba_{0.84}K_{0.16}Fe_2As_2$.

The lower bulk $T_c$ values in the pressure range of the re-entrant tetragonal phase demonstrate that the re-entrant $C_4$ phase competes particularly strong with superconductivity under pressure. The onset of filamentary superconductivity associated with the orthorhombic minority domains can be taken as estimation how high $T_c$ could be if the re-entrant $C_4$ phase would be absent. The specific heat data, which provides the bulk thermodynamic values of the transition temperatures, reveals that the bulk $T_c$ is suppressed as much as ~20 K (from 34 K down to 13.5 K) by the re-entrant tetragonal order.

## ACKNOWLEDGMENTS

R.L. will always be grateful to D. Jaccard for sharing his extraordinary expertise in high-pressure experiments with him. We further thank U. Lampe for technical support. This work was supported by grants from the Research Grants Council of the Hong Kong Special Administrative Region, China (603010, SEG_HKUST03, SRFI11SC02).